\documentclass[twocolumn,twoside,amsmath,amssymb,floatfix]{revtex4}

\usepackage{pslatex}
\usepackage{epsfig}
\usepackage{graphicx}
\usepackage{dcolumn}
\usepackage{bm}

\begin{document}
 
\preprint{APL/123-QED}

\title{A conjecture on a quantum limit of semiconductor lasers}
\author{P. Harrison and E. E. Orlova}
\affiliation{Institute of Microwaves and Photonics, 
School of Electronic and Electrical Engineering, University of Leeds, LS2 9JT,
United Kingdom}

\date{\today}

\begin{abstract}
A relationship between the maximum operating temperature of semiconductor
lasers and their emission wavelength is conjectured.  The conjecture is
supported by a wide variety of existing experimental data for visible and 
infrared double
heterojunction and quantum well lasers, quantum cascade lasers, as well as 
more esoteric devices such as Type-II antimonide-based mid-infrared and
$p$-Ge and impurity-based Terahertz devices.  The relationship developed 
may enable the ultimate performance of mid- and far-infrared, as well as
Terahertz semiconductor lasers to be predicted.
\end{abstract}


\maketitle

The Terahertz region of the spectrum\cite{miles01b} remained the last
region of the electromagnetic spectrum to be explored for one simple
reason: it lies at the junction between the electronic technologies of
increasingly higher frequency current (electron) oscillations in
millimetre wave devices and the longer wavelength transitions between
discrete states which characterise infrared optical technologies.  
Subsequently the Terahertz region of the spectrum lies at the 
boundary between classical and quantum physics.

Semiconductor light emitting diodes and lasers produce quanta of
electromagnetic radiation (photons) of energy $h\nu$ through the
transitions of electrons between states, where $h$ is Planck's constant and
$\nu$ is the frequency of the electromagnetic radiation.  
In quantum dot\cite{bimberg99b}
and impurity-based devices\cite{hubers99p2655} these states are discrete, 
however in the vast majority of devices these quantised states broaden into 
bands---a continuous range of energies between two fixed
limits.  In contrast to gas lasers and masers which can produce microwave 
to optical frequencies at room temperature, solid state lasers have the 
parallel mechanism of non-radiative phonon emission (lattice vibrations)
competing against photon 
(light) emission.  Again it is another point of interest that Terahertz 
represents the range of frequencies at which the photon energies $h\nu$
are of the same order as the phonon energies $\hbar\omega$
($\hbar=h/(2\pi)$ and $\omega$ is the angular frequency of the lattice
vibration).  The competing
non-radiative phonon emission rate is dependent on $N_0+1$,
where the phonon occupation number $N_0$ is given by:
\begin{equation}
N_0=\frac{1}{\exp{(\hbar\omega/kT)}-1}
\label{eq:N0}
\end{equation}
with $kT$ representing the thermal energy density within the crystal lattice.

\begin{figure}[htb]
\epsfig{figure=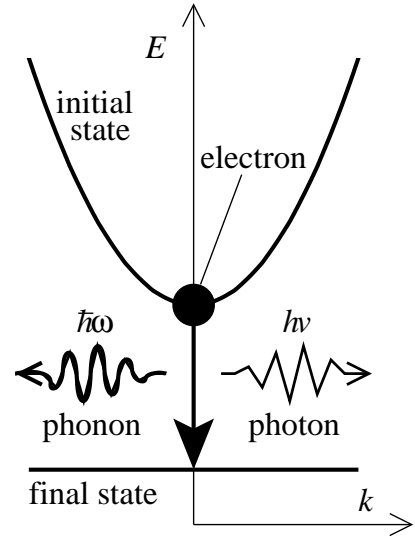,width=0.3\textwidth}
\caption{Schematic diagram illustrating the competition between radiative
(photon-generating $h\nu$) and non-radiative (phonon-generating $\hbar\omega$)
transitions.}
\label{fig:fig1}
\end{figure}

Until recently there has been no need to try and link the energies
$h\nu$, $\hbar\omega$ and $kT$ together, but the impetus driving the 
development of mid-infrared\cite{gmachl01p1533} and Terahertz quantum cascade 
lasers\cite{kohler02p156} has begun to bridge the gap between the 
classical and quantum worlds and the questions \emph{`What is the ultimate long
wavelength limit?'} and \emph{`What are the maximum operating temperatures 
of Terahertz quantum cascade lasers?'} have started to arise more frequently.

There are complicated approaches which could be followed to try to answer
these questions, for example, one could use a physical model of quantum
cascade lasers\cite{harrison01p153,indjin02p9019} and could do lengthy
calculations of the gain versus current profile for designs for a large
number of wavelengths.  Alternatively one could resort to experimental data 
and derive an empirical relationship.  But neither of these methods is
quick, the results would not be that transparent and doubts would remain
over generality to different material systems and different device designs.

\begin{figure}[htb]
\epsfig{figure=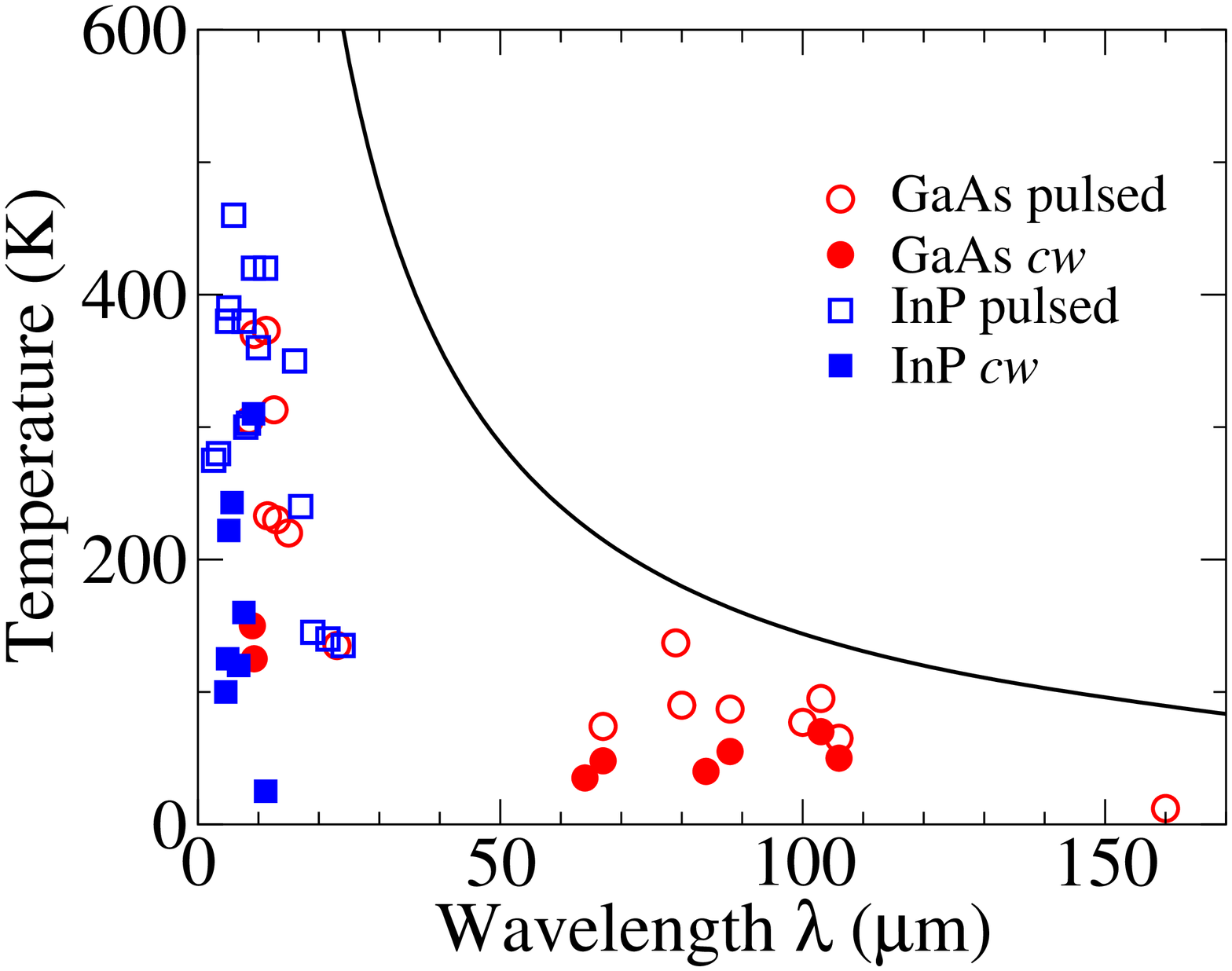,width=0.4\textwidth}
\caption{The points represent the highest operating temperature as a
function of emission wavelength for state-of-the-art quantum cascade 
lasers as reported in the
literature\cite{carder03p3409,page02p1312,page03ITQW,anders02p1864,gianordoli00p1144,page01p556,ulrich02p3691,barbieri04APL,williams03p5142,ajili02p1675,williams03p2124,williams03p915,tredicucci03ITQW,scalari03ITQW,kohler04p1266,scalari03p3165,barbieri03p586,kohler03p1518,faist98p680,rochat01p4271,tredicucci99p638,tredicucci00p2164,tredicucci00p2286,colombelli01p2620,hofstetter01p396,soibel03p24,hofstetter01p1964,yu03p3397,yarekha03p1123,blaser03ITQW,mann03ITQW,schrenk03ITQW,faist02p533,capasso02p511}.  The solid line
represents the speculated `quantum limit' $h\nu=kT$.}
\label{fig:fig2}
\end{figure}

It is clear that when the phonon energy $\hbar\omega$ is equal to the energy
separation between the quantised energy levels (and hence the photon energy
$h\nu$) that it might be expected that the detrimental phonon emission
process will begin to be significant.  However, at very
low temperatures ($kT\ll\hbar\omega$, see equation~\ref{eq:N0}) there are 
few phonons, with these processes perhaps becoming significant when the
thermal energy is of the order of the phonon energy, i.e. $kT=\hbar\omega$,
see Fig.~\ref{fig:fig1}.
Thus in the search for a relationship between the emission wavelength $\lambda$
of a semiconductor laser and its maximum operating temperature $T$, one is
guided by two arguments to suggest that favourable conditions for
sustaining a population inversion and achieving lasing occur when:
\begin{equation}
h\nu>\hbar\omega\hspace*{5mm}\text{and}\hspace*{5mm}kT<\hbar\omega
\end{equation}
which can be combined to give:
\begin{equation}
h\nu>kT\hspace*{5mm}\text{or}\hspace*{5mm}
\frac{hc}{\lambda}>kT
\label{eq:hnu}
\end{equation}
where $c$ is the speed of light.
This relationship looks too simple, so it is important to look around for any
supporting evidence and this is provided by the data in Fig.~\ref{fig:fig2}
which plots the maximum operating temperature against
wavelength for state-of-the art quantum cascade lasers in both the
InGaAs/AlInAs on InP and the GaAs/AlGaAs on GaAs systems for pulsed and
continuous wave (\emph{cw}) operation.  The figure also shows the
`limit' predicted by equation~\ref{eq:hnu} as a solid line.  The
surprising point to note is that all experimental data to date obeys the
limit.
  
There is nothing special about quantum cascade lasers and so it is
worthwhile seeing if experimental data from other types of semiconductor
laser conform to this speculated limit.  The type-II interband cascade
lasers have operating wavelengths in the short-wave (3--5 $\mu$m) infrared
and are now approaching room
temperature\cite{bewley99p1075,yang99p1254,lee99p1743} and the lead salt
lasers have emission wavelengths from 3--20 $\mu$m and $cw$ operation above
200 K\cite{tacke01p547} which puts both of these categories below the solid
line and in the bottom left hand corner of Fig.~\ref{fig:fig2}.  Looking back 
at Fig.~\ref{fig:fig2} is is clear that it is in the Terahertz region of the
spectrum where the quantum cascade lasers come closest to reaching the
limit of equation~\ref{eq:hnu}, so it is important to look at other types of
Terahertz semiconductor laser.  Br\"undermann \emph{et
al.}\cite{brundermann98p2757} report on a range of $p$-Ge Terahertz lasers 
which can be tuned from 1 to 4 THz (300--75 $\mu$m).
In these devices the holes are accelerated by an electric field under the
influence of an external magnetic field, Br\"undermann reports maximum 
operating temperatures up to 36 K and hence these devices also obey the 
limit (Note at a wavelength of 300 $\mu$m, equation~\ref{eq:hnu} would 
imply a temperature of 47 K).  It is this class of devices which is likely
to challenge the limit most strongly with 
Komiyama \emph{et al.}\cite{komiyama91pS133}
reporting emission at 1.8 mm at 4.2 K (equation~\ref{eq:hnu} suggests a
maximum operating temperature for lasing as 7 K).  Even quantum-dot-like 
devices such as those which use internal impurity transitions to generate 
Terahertz radiation, see for example \cite{hubers02p191,pavlov03p126}, are 
limited to low temperature operation.  In the same vein, the application of
a magnetic field along the growth axis of semiconductor heterostructures
leads to the in-plane localisation of charge carriers and hence the
discretisation of the quantum well subbands.  In the case of quantum
cascade lasers this has lead to the so-called `quantum box cascade
lasers'\cite{becker02p2941}, but investigations of these devices have been
focussed on reducing the threshold current or increasing the luminescence
intensity\cite{alton03a081303r,blaser02p67,tamosiunas03p3873} or inducing
lasing that doesn't otherwise exist\cite{scalari03p3453} and not on
increasing the operating temperature or wavelength.

In summary, a simple relationship which links the maximum operating 
temperature of a semiconductor laser with its emission wavelength has 
been proposed.  A great deal of experimental data has been cited which
supports this `quantum limit', however as the limit has been argued
and not rigorously proven, it can only have the status of a conjecture which
it is hoped will stimulate discussion.

PH would like to thank D. Indjin for collecting the QCL data.


\bibliography{main,ph}


%
%

\end{document}